\documentclass[3p,times]{elsarticle}

\usepackage{ecrc}

\usepackage{enumitem}  % Load the enumitem package for list customization
\setlist[enumerate]{itemsep=0pt, topsep=0pt}
\usepackage{url}

\volume{00}

\firstpage{1}

\journalname{Astronomy and Computing}

\runauth{T. de Haan}

\jid{Astron. Comput.}

\jnltitlelogo{Astronomy and Computing}

\usepackage{amssymb}
\usepackage{lineno}

\usepackage[figuresright]{rotating}

\usepackage{xcolor}
\newcommand{\reply}[1]{\textcolor{black}{#1}}
\newcommand{\secondreply}[1]{\textcolor{black}{#1}}
\newcommand{\thirdreply}[1]{\textcolor{black}{#1}}

\usepackage{mdframed}
\newcounter{box}

\begin{document}

\begin{frontmatter}



\dochead{}
%% e.g. \dochead{17th International Conference on Dynamical Processes in Excited States of Solids}

\title{\textsc{cosmosage}: A Natural-Language Assistant for Cosmology}
\thispagestyle{empty}

\author[1,2]{Tijmen de Haan}
\ead{tijmen.dehaan@gmail.com}
\address[1]{Institute of Particle and Nuclear Studies (IPNS), High Energy Accelerator Research Organization (KEK), Tsukuba, Ibaraki 305-0801, Japan}
\address[2]{International Center for Quantum-field Measurement Systems for Studies of the Universe and Particles (QUP-WPI), High Energy Accelerator Research Organization (KEK), Tsukuba, Ibaraki 305-0801, Japan}

\begin{abstract}
\textsc{cosmosage} is a natural-language assistant intended for a wide audience, from laypersons interested in cosmology to students, teachers, and professional cosmologists. \textsc{cosmosage} provides a novel way to access knowledge and reason about cosmology. Leveraging the power of advanced large language models (LLMs), \textsc{cosmosage} has learned from a vast corpus of open-access source texts, including textbooks and papers. \textsc{cosmosage} is found to be state-of-the-art on the narrow task of answering questions about cosmology, outperforming all general-purpose models. The model parameters and code are publicly available.
\end{abstract}


\end{frontmatter}

%\linenumbers


\section{Introduction}

Large language models (LLMs) are a recent and exciting development in artificial intelligence that have shown remarkable capabilities in understanding and generating human-like text. Powered by a decoder-only version of the transformer architecture \citep{vaswani_attention_2017, brown_language_2020}, \mbox{OpenAI}'s GPT-3 caught the public's imagination with the launch of ChatGPT in November 2022. \reply{By training on a significant fraction of the text data available on the internet, LLMs have learned to understand and produce language with human-like fluency and coherence.  While the exact nature of their capabilities remains a subject of active research \citep{chollet_measure_2019, mondorf_beyond_2024, li_emergent_2024}, LLMs have demonstrated practical utility in addressing novel problems and tasks. This opens up new possibilities for natural language interfaces that can assist humans in a wide range of tasks, including personalized teaching, aiding in idea generation, as coding assistants, and for creative and/or scientific writing tasks.}

In this work, I show that a relatively small model that is narrowly tailored to a particular application outperforms state-of-the-art models with general capabilities. The model is named \textsc{cosmosage}, an LLM that is fine-tuned from publicly available base models such as \textsc{Mistral-7B-v0.1} \citep{jiang_mistral_2023}, a relatively modest-sized LLM base model with 7.24 billion learned parameters. After continued pretraining in the form of text completion on the data corpus, I performed assistant-tuning to create a natural-language assistant specifically tailored for answering questions about the field of cosmology. By training on a large body of open-access cosmology textbooks, papers, and other resources, \textsc{cosmosage} has acquired a deep understanding of cosmological concepts, theories, and research.

\section{Development of \protect{\textsc{cosmosage}}}

\begin{figure}
\centering
\includegraphics[width=\linewidth]{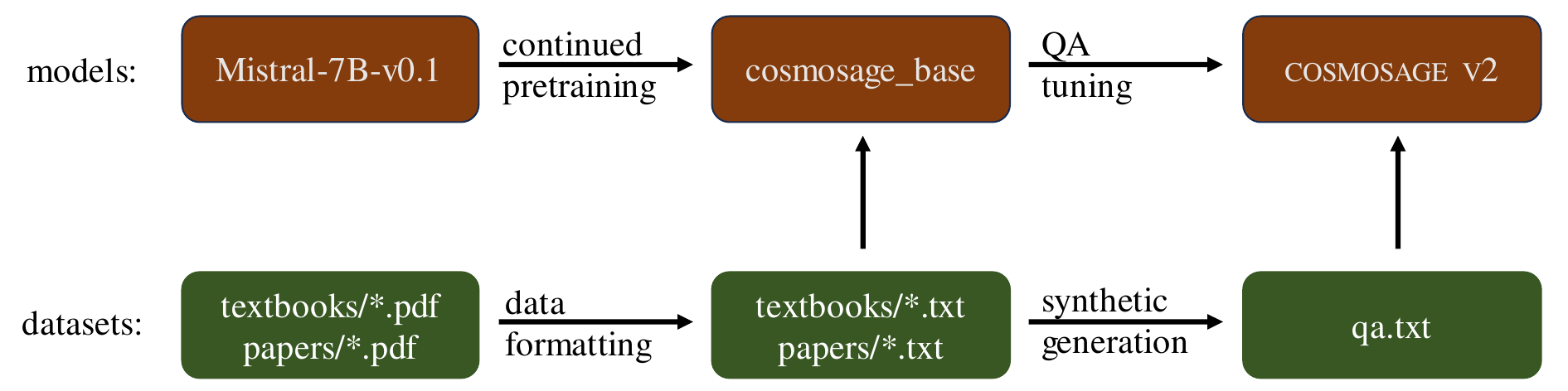}
\caption{Overview of the training procedure for \textsc{cosmosage~v2}. The initial continued pretraining adds knowledge to the base model. The subsequent tuning on question-answer pairs creates a useful assistant model.}
\label{fig:flowchart}
\end{figure}

The development of \textsc{cosmosage} involved several steps, summarized in Figure~\ref{fig:flowchart}.

\subsection{Foundation Model}

\secondreply{
The \textsc{cosmosage} project has undergone a number of releases which differ primarily by the choice of starting foundation model, also known as a base model. At the time of writing, nine versions of \textsc{cosmosage} are available. This article focuses on three models: \textsc{cosmosage~v2}, based on \textsc{Mistral-7B-v0.1} \citep{jiang_mistral_2023}, \textsc{cosmosage~v3}, based on \textsc{META-LLAMA-3-8B}, and \textsc{cosmosage~v3.1}, based on \textsc{META-LLAMA-3.1-8B} \citep{dubey_llama_2024}.
}

\secondreply{The \textsc{Mistral} model, like other modern language models, is built on the transformer architecture \citep{vaswani_attention_2017}. At its core, the transformer processes text as a sequence of tokens, each representing a word or subword unit. This sequence is first embedded into a 4096-dimensional space. The model’s 32 transformer layers each incorporate self-attention mechanisms to capture the relationships between tokens. These layers are organized into 32 parallel heads, each processing different aspects of the input. After these 32 heads, a dense layer maps the resulting representations to the final output: a probability distribution over the next token in the sequence.}

\secondreply{At inference time, the model is run autoregressively. The sequence containing the system prompt and user prompt is presented, as well as the header for a blank assistant response. The model is then run, and the next token is sampled from the probability distribution using some sampling rules. The model is the run over the new sequence and another token is sampled. This process is repeated until the new assistant response is completed, indicated by a special end-of-sequence token.}

\subsection{Data Collection and Preprocessing}

The source documents for \textsc{cosmosage} consist of textbooks, lecture notes, papers, and other open-access datasets. The textbooks were found manually by searching the internet. The papers were selected using a mixture of search queries and preference of the cosmic microwave background community as determined by upvotes on my website that hosts a modified version of \texttt{arxiv-sanity-lite} \citep{karpathy_arxiv-sanity-lite_2024}. All papers were downloaded with the arXiv API. Finally, all non-PDF documents, such as textbooks in non-PDF eBook formats were converted to PDF using command line scripts. 

The PDFs were then converted to human-readable lightweight markdown format using Nougat OCR \citep{blecher_nougat_2023}. This markdown text required some cleaning by removing repeated words and repeated sequences via a recursive algorithm. Furthermore, malformatted outputs were spotted by eye and for each of these occurrences an appropriate regular expression was created for search-and-delete. This process was continued until no further malformatted markdown was spotted, requiring a total of 10 regular expressions. 

\begin{figure}
\centering
\includegraphics[width=0.4\linewidth]{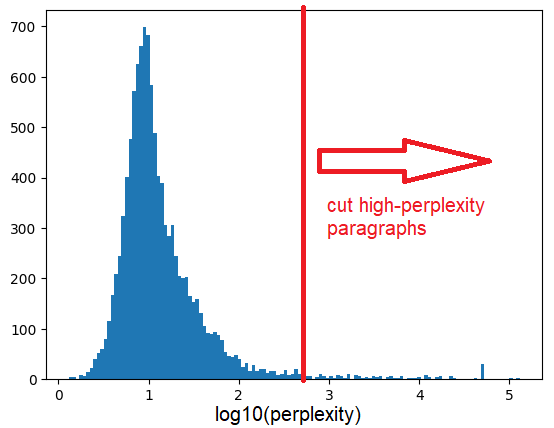}
\caption{Histogram of perplexity over paragraphs. The lowest perplexity paragraphs were excised from the continued pretraining dataset.}
\label{fig:perplexity}
\end{figure}

One further cleaning step was performed starting with \textsc{cosmosage~v3.1}, where the data was split into paragraphs and a forward pass was done with the base model, storing the average loss for the paragraph. The perplexity of that paragraph is defined as $\exp(-\left< \mathrm{loss} \right>)$. The perplexity histogram is shown in Figure~\ref{fig:perplexity}. The top 2.7\% of high-perplexity paragraphs were deleted. No documents were left without any paragraphs.

\subsection{Synthetic Question-Answer Pair Generation}
\label{sec:synthqa}

The second stage of training---assistant tuning---requires triplets of a system prompt, user prompt, and assistant completion. The assistant completion is the target over which the loss of each token is calculated given the previous tokens. Of paramount importance are quality, quantity, and diversity of the dataset. The following procedure was employed.

First, the input data was separated into overlapping segments. The segment length was set to 1800 characters, with 600 characters of overlap. The 8-bit GPTQ version of the \verb||{bagel-34b-v0.2} \citep{durbin_bagel-34b-v02_2023} model was used with the following prompt. The system prompt was given as
\begin{quote}
\textit{Assume the role of a cosmology lecturer with a deep understanding of all areas of modern cosmology, astrophysics. You must precisely follow the instructions given by the user.}
\end{quote}
and the user prompt:
\begin{quote}
\textit{As a cosmology expert, your task is to create precise and self-contained question-answer pairs from a specified PASSAGE of} \{shorthand\_title\}.
\textit{The paper is titled} \{title\} \textit{and has the arXiv ID} \{arxiv\_id\}. \textit{Here is a short summary:}
\{summary\}
\textit{Ensure that each question incorporates all necessary context, allowing it to be fully understood on its own. 
Answers should be clear, specific, and provide comprehensive information based on the PASSAGE. 
The goal is for each question and answer pair to be understandable independently, ensuring they are complete and contextually clear without external references.}

\textit{PASSAGE:}
\{input text segment\}
\end{quote}
In order to encourage variety, an additional instruction was inserted into the user prompt prior to the ``PASSAGE'', randomly selected from 20 manually crafted additional instructions. These additional instructions include phrases such as ``\textit{Consider whether a mathematical representation would be helpful in summarizing the PASSAGE.}'' and ``\textit{Create questions based on hypothetical scenarios or 'what-if' questions inspired by the PASSAGE.}''

To enhance the quality of the question-answer (QA) pairs, we implemented a refinement procedure. This process was designed to mimic an expert teacher evaluating and improving a student's work, with the ultimate goal of creating high-quality synthetic QA pairs. Given the well-documented tendency of Large Language Models (LLMs) to overestimate the quality of their own outputs \citep{lin_truthfulqa_2022}, we employed the \textsc{Mixtral-8x7B-v0.1-instruct} model \citep{jiang_mixtral_2024} as an independent evaluator. 

\noindent The refinement procedure was structured as follows:
\begin{enumerate}
    \item Each QA pair was presented to the \textsc{Mixtral} model along with the original context used to generate the pair.
    \item The model was instructed to assign a numerical grade (0-100\%) to the answer based on its accuracy, completeness, and relevance to the question.
    \item If the original answer received a grade of 90\% or higher, it was retained in the dataset.
    \item If the original answer scored below 90\%, the \textsc{Mixtral} model was prompted to generate an improved ``solution key'' answer.
    \item This solution key was then evaluated using the same grading criteria.
    \item If the solution key received a grade of 90\% or higher, it replaced the original answer in the QA pair.
    \item If both the original answer and the solution key scored below 90\%, the entire QA pair was discarded.
\end{enumerate}

\noindent This iterative refinement process resulted in a curated dataset of over 1.3 million high-quality QA pairs. The combination of the initial generation and refinement used approximately 500 A100 GPU-hours.

\subsection{Model Training}

The initial step in further training was to continue pretraining this base model on our corpus of cosmology texts. While early versions of \textsc{cosmosage} used a custom pytorch training loop, the \texttt{axolotl} \citep{openaccess-ai_collective_openaccess-ai-collectiveaxolotl_2024} training framework was found to be preferable given its large number of implemented algorithms and ease of switching between these. The AdamW optimizer \citep{kingma_adam_2017, loshchilov_decoupled_2019} was used with the \texttt{deepspeed\_zero\_stage\_2} configuration provided by the DeepSpeed library \citep{rajbhandari_zero_2020}. The learning rate was set to $1\times10^{-5}$ with a micro batch size of 1, over 8 GPUs. The training was performed for three epochs, taking approximately 180 A100 GPU hours.

Following the continued pretraining phase, we implemented a fine-tuning process to enhance \textsc{cosmosage}'s capabilities in natural language interaction and cosmology-specific question answering. This phase used supervised learning on a diverse dataset of question-answer pairs and dialogue examples, using the checkpoint obtained from the continued pretraining as the starting point.

The fine-tuning dataset comprised the refined cosmology-specific QA pairs generated in Section~\ref{sec:synthqa}, as well as procedurally generated QA pairs about metadata, and QA pairs based on synthetically generated summaries of the cosmology papers. To enhance the model's ability to engage in multi-turn conversations and follow general instructions, we also incorporated the \texttt{OpenHermes 2.5} dataset \citep{teknium_openhermes_2023}.

The fine-tuning process employed the same \texttt{axolotl} framework and DeepSpeed configuration as the continued pretraining phase, ensuring consistency in the training pipeline. However, we adjusted the hyperparameters to suit the fine-tuning task by decreasing the initial learning rate to $2 \times 10^{-6}$, adding a linear decay with warmup over 100 steps. The fine-tuning process was performed over two epochs, using approximately 300 A100 GPU hours. 

\section{Evaluation}

\reply{The capabilities of \textsc{cosmosage} were primarily assessed in three primary ways: subjective qualitative evaluation through ad-hoc interaction, quantitative evaluation with non-cosmology benchmarks, and quantitative blind preference testing.}

\subsection{Qualitative Evaluation}

In the first category of qualitative evaluation through practical use, the following three examples illustrate instances where \textsc{cosmosage~v3} demonstrates superior performance compared to state-of-the-art generalist models when answering cosmology-related questions. However, it is important to note that as a relatively small language model, \textsc{cosmosage}'s reasoning capabilities are not as strong as those of very large models such as Claude or ChatGPT. For questions that demand extended logical reasoning or precise adherence to complex instructions, larger models remain preferable.

The first example is a comparison of \textsc{cosmosage} and Claude Sonnet 3.5 (2024-06-20 version) on a cosmology question testing model knowledge. \textsc{cosmosage} consistently outperforms large generalist models on questions of this nature.

\begin{mdframed}
\small
\textbf{Question:} How many parameters are in the wCDM model?

\textbf{\textsc{cosmosage} answer:} The wCDM model is a cosmological model that includes seven parameters: $\Omega_\mathrm{b} h^2$, $\Omega_\mathrm{c} h^2$, $100 \theta_\mathrm{MC}$, $\tau_\mathrm{reio}$, $n_\mathrm{s}$, $\ln(10^{10} A_\mathrm{s})$, and $w$. These parameters describe the composition and evolution of the universe. Let's break down each parameter:
\begin{enumerate}
    \item $\Omega_\mathrm{b} h^2$: This parameter represents the density of baryonic matter (normal matter) in the universe, multiplied by the square of the Hubble constant ($H_\mathrm{0}$). It is expressed as a fraction of the critical density.
    \item $\Omega_\mathrm{c} h^2$: This parameter represents the density of cold dark matter in the universe, multiplied by the square of the Hubble constant ($H_\mathrm{0}$). It is also expressed as a fraction of the critical density.
    \item $100 \theta_\mathrm{MC}$: This parameter is related to the angular size of the sound horizon at recombination. The sound horizon is the distance that sound waves could travel before the universe became transparent to radiation. This value is used to determine the geometry of the universe.
    \item $\tau_\mathrm{reio}$: This parameter represents the optical depth due to reionization, which is the process where neutral hydrogen atoms in the early universe were ionized by photons from hot stars. It quantifies the amount of scattering of photons during this period.
    \item $n_\mathrm{s}$: This parameter represents the spectral index of primordial scalar perturbations. It describes the distribution of density fluctuations in the early universe, which eventually led to the formation of galaxies and other large-scale structures.
    \item $\ln(10^{10} A_\mathrm{s})$: This parameter represents the amplitude of the initial power spectrum of density fluctuations. It is a measure of the strength of these fluctuations, which also contributed to the formation of galaxies and other large-scale structures.
    \item $w$: This parameter represents the equation of state for dark energy. Dark energy is a mysterious component that drives the accelerated expansion of the universe. By allowing $w$ to vary, the wCDM model can account for different models of dark energy, such as quintessence or phantom energy.
\end{enumerate}
In summary, the wCDM model has seven parameters that describe the composition, geometry, and evolution of the universe, including the properties of matter, dark energy, and the initial conditions of the Big Bang.

\textbf{Claude answer:} The wCDM model, which is an extension of the standard $\Lambda$CDM (Lambda Cold Dark Matter) model, typically has 6 main parameters:

\begin{enumerate}
    \item $\Omega_\mathrm{m}$ ($\Omega_\mathrm{m}$): The density parameter for matter (including both baryonic and dark matter).
    \item $\Omega_\mathrm{b}$ ($\Omega_\mathrm{b}$): The density parameter for baryonic matter.
    \item $h$: The Hubble parameter ($H_\mathrm{0} = 100h$ km/s/Mpc).
    \item $n_\mathrm{s}$: The spectral index of primordial density fluctuations.
    \item $\sigma_8$ ($\sigma_8$): The amplitude of matter fluctuations on $8 \, \mathrm{Mpc}/h$ scales.
    \item $w$: The dark energy equation of state parameter.
\end{enumerate}

The key difference between $\Lambda$CDM and wCDM is the inclusion of $w$ as a free parameter. In $\Lambda$CDM, $w$ is fixed at -1, corresponding to a cosmological constant. In wCDM, $w$ is allowed to vary, potentially differing from -1.

It's worth noting that depending on the specific analysis or dataset being used, additional parameters might be included, such as:

\begin{itemize}
    \item $\tau$ ($\tau$): The optical depth to reionization.
    \item $\Omega_\mathrm{k}$ ($\Omega_\mathrm{k}$): The curvature parameter.
    \item $N_\mathrm{eff}$: The effective number of relativistic species.
\end{itemize}

However, the core wCDM model is typically described by the 6 parameters listed above. The exact number of parameters can vary slightly depending on the specific formulation or application of the model.
\end{mdframed}

%The second example shows a comparison of responses to an ambiguous query between \textsc{cosmosage} and ChatGPT (GPT-4o, 2024-07-03). \textsc{cosmosage}, fine-tuned for cosmology, correctly interprets the context and provides a relevant explanation of the term ``stacking.''

\reply{The second example shows how \textsc{cosmosage} handles ambiguous or underspecified queries---a common scenario in real-world interactions where users may ask brief or imprecise questions without providing full context. The term ``stacking'' has different meanings in different fields---in machine learning it refers to combining multiple base models to improve prediction accuracy, while in astronomy it refers to the co-addition of multiple observations so that the signal adds coherently, but the noise adds incoherently, increasing overall signal-to-noise ratio. \textsc{cosmosage}, being specialized in cosmology and astronomy, automatically interprets such queries within the astronomical context without requiring explicit disambiguation. This ``cosmology-first'' approach is particularly valuable for students, researchers, and enthusiasts who want direct answers to their cosmology questions without having to carefully phrase queries to avoid ambiguity with other fields. In contrast, ChatGPT defaults to the machine learning interpretation, demonstrating how domain-specific training helps \textsc{cosmosage} provide more relevant responses even to casually worded queries.}

\begin{mdframed}
\small
\textbf{Question:} What is "stacking"?

\textbf{\textsc{cosmosage} answer:} Stacking is a technique used in astronomy to combine multiple images or data sets of the same object into a single image or dataset. This can be done by aligning and co-adding the individual images or data points, which helps to improve the signal-to-noise ratio and reveal faint features that may not be visible in a single image or dataset. Stacking is often used to study the properties of galaxies, galaxy clusters, and other astronomical objects.

\textbf{ChatGPT answer:} Stacking is an ensemble learning technique used in machine learning to improve the performance of predictive models. It involves combining multiple different models (often referred to as base models or level-0 models) to create a more robust and accurate predictive model.

Here’s how stacking generally works:

\begin{itemize}
    \item \textbf{Train Base Models:} Multiple base models are trained on the same dataset. These models can be of different types (e.g., decision trees, logistic regression, neural networks) or variations of the same type with different hyperparameters.
    \item \textbf{Generate Meta-Features:} Each base model makes predictions on the training data. These predictions, along with the original features, 
\end{itemize}
[response continues]
\end{mdframed}

The final example is one where \textsc{cosmosage} is asked to do something it is not capable of doing: a non-trivial calculation that requires numerical integration. Despite this, it makes a valiant attempt by using the Hubble-Lemaître law (which is approximately correct for low-redshift objects) and gets close to the correct answer of 918.4~Mpc, despite inaccurate arithmetic division. The completions for Claude Sonnet 3.5, GPT-4o, LLAMA-70B and Phi-3 are not shown for brevity, but all perform erroneous calculations, often attempting the numerical integral and hallucinating the result.

\begin{mdframed}
\small

\textbf{Question:} Calculate the angular diameter distance to the Bullet cluster. Show your work.

\textbf{\textsc{cosmosage} answer:} To calculate the angular diameter distance (dA) to the Bullet cluster, we need to know its redshift (z). The redshift of the Bullet cluster is approximately 0.296.

The angular diameter distance is related to the comoving distance by the formula:

dA(z) = dC(z)/(1 + z)

where dA(z) is the angular diameter distance, dC(z) is the comoving distance, and z is the redshift. 

First, let's find the comoving distance using the Hubble constant (H0), which is approximately 70 km/s/Mpc. We can use the Hubble-Lemaître law for the comoving distance:

dC(z) $\approx$ c/H0 * z

where c is the speed of light (approximately 300,000 km/s).

Now, let's convert the redshift from dimensionless to a proper length:

z $\approx$ 0.296
c * z $\approx$ 89,200 km/s

Now, we can find the comoving distance:

dC(z) $\approx$ 89,200 km/s / 70 km/s/Mpc $\approx$ 1273 Mpc

Finally, we can find the angular diameter distance:

dA(z) = dC(z)/(1 + z)
dA(z) = 1273 Mpc / (1 + 0.296)
dA(z) $\approx$ 1035 Mpc

So, the angular diameter distance to the Bullet cluster is approximately 1035 Mpc.
\end{mdframed}

\reply{These three examples highlight a few aspects of \textsc{cosmosage}'s behavior. It is generally strong at knowledge-based questions, it assumes a cosmology context, and it is relatively weak at tasks requiring deep reasoning such as calculations.}

A common concern when presenting a new model is that the examples shown are carefully selected, or ``cherry-picked,'' to highlight the model's strengths and downplay its weaknesses. To avoid such concerns, here is the explicit way in which the examples showcased in this work were selected. 

The generalist models used for comparison (ChatGPT and Claude) were chosen due to their community perception as the strongest generalist models at the time of writing. Each model is shown once and this selection was based on the particular output. \reply{The \textsc{cosmosage} outputs presented above were generated only once, except for the calculation example which was generated twice due to an incorrect initial response.}

\subsection{General-purpose Benchmarks}

\reply{\textsc{cosmosage} was submitted to the publicly available benchmark suite ``Open LLM Leaderboard'', which evaluates models on six diverse tasks. The AI2 Reasoning Challenge (ARC) \cite{clark_think_2018} tests scientific reasoning through multiple-choice questions. HellaSwag \cite{zellers_hellaswag_2019} evaluates commonsense inference by having models complete scenarios with plausible endings. MMLU (Massive Multitask Language Understanding) \cite{hendrycks_measuring_2021} spans 57 subjects including science, humanities, and professional knowledge. TruthfulQA \cite{lin_truthfulqa_2022} measures a model's ability to generate truthful rather than false statements. Winogrande \cite{sakaguchi_winogrande_2019} tests common-sense reasoning. GSM8k \cite{cobbe_training_2021} valuates mathematical problem-solving through grade school word problems. All scores are reported as accuracy percentages, representing the proportion of correct responses out of the total number of test questions. The ``Avg.'' score is the arithmetic mean across all six individual test scores. The results are shown in Table \ref{tab:openllmleaderboard}, showing that the specialization in cosmology only slightly reduced the general capabilities of the model, and in fact has enhanced the capability of the model on the grade-school mathematics dataset GSM-8k.}

\begin{table}
\centering
\begin{tabular}{|l|c|c|}

\hline
\textbf{Evaluation} & \textsc{cosmosage~v2} & \texttt{OpenHermes2.5-Mistral7B} \\ \hline
Avg.                             & 60.66 & 61.52 \\ \hline
AI2 Reasoning Challenge (25-Shot) & 59.73 & 64.93 \\ \hline
HellaSwag (10-Shot)              & 80.90 & 84.18 \\ \hline
MMLU (5-Shot)                    & 59.57 & 63.64 \\ \hline
TruthfulQA (0-shot)              & 50.98 & 52.24 \\ \hline
Winogrande (5-shot)              & 75.93 & 78.06 \\ \hline
GSM8k (5-shot)                   & 36.85 & 26.08 \\ \hline

\end{tabular}
\caption{Performance comparison on Open LLM Leaderboard benchmarks for non-cosmology skills. \textsc{cosmosage} and \texttt{OpenHermes2.5-Mistral7B} share a base model and instruction-tuning dataset, differing only in \textsc{cosmosage}'s additional cosmology-specific training data, as well as potential differences in training hyperparameters. Results suggest that cosmology specialization did not substantially impact performance on general tasks.}
\label{tab:openllmleaderboard}
\end{table}

\subsection{Independent Blind Human Ranking}

\thirdreply{To quantitatively assess \textsc{cosmosage}'s performance on cosmology-specific Q\&A, a blind comparative evaluation was conducted using fifteen questions spanning diverse areas of cosmology, ranging from layperson to professional cosmologist-level difficulty. Independent evaluators were asked to compare responses from \textsc{cosmosage~v3.1} against those from \textsc{META-LLAMA-3.1-8B-Instruct}, a general-purpose model trained by Meta on their proprietary dataset using far more computational resources than \textsc{cosmosage}. Both models received identical system prompts explaining the cosmology context.}

\thirdreply{Responses were presented in randomized order, with evaluators asked to grade the overall quality of each response while remaining unaware of their source. Three evaluators participated in the blind ranking process. The evaluators showed remarkable agreement in their assessments, with unanimous agreement on all but one question. In this case, \textsc{META-LLAMA-3.1-8B-Instruct} provided a longer, more complete answer that contained a hallucination, while \textsc{cosmosage}'s response was very terse but accurate. This high level of agreement (14 out of 15) suggests a clear and consistent difference in quality between the two models' responses. After unblinding, \textsc{cosmosage}'s responses were preferred 75.6\% of the time, demonstrating a highly statistically significant preference ($p < 2 \times 10^{-4}$) for \textsc{cosmosage~v3.1} over \textsc{META-LLAMA-3.1-8B-Instruct}.}

\secondreply{This result shows that the narrowly focused post-training on cosmology-related datasets yields superior performance to the more thorough but less focused training on general-purpose datasets performed by Meta.}

\section{Availability of Products}

The \textsc{cosmosage} project aims to provide a useful natural language assistant to the cosmology community and anyone interested in cosmology. Therefore, to promote openness, collaboration, and visibility, various components have been made publicly available under the extremely permissive MIT license. The source code used for training and inference, along with documentation and usage examples, is publicly available on GitHub at:
\\ \ \url{https://github.com/tijmen/cosmosage} 

For researchers and developers interested in experimenting with the model directly, the trained \textsc{cosmosage} model parameters (``weights'') are hosted on Hugging Face at
\\ \ \url{https://huggingface.co/Tijmen2/} 

Furthermore, despite the resource-intensive nature of running large language models, \textsc{cosmosage} has been deployed for academic use on the Helmholtz AI Blablador inference server hosted by Jülich Supercomputing Centre (JSC): 
\\ \ \url{https://helmholtz-blablador.fz-juelich.de/}

\section{Potential Enhancements and Extensions}

While \textsc{cosmosage} shows significant capabilities in its current state, there are numerous avenues for improvement. A clear next step is to rerun the training procedure on an larger, stronger, and newer base model, such as \texttt{gemma-2-27b}. 

Another direction that I am in the process of exploring is to give the model access to the source material at inference time via retrieval augmented generation (RAG). This technique allows the LLM to access and utilize the original material, reducing the frequency of hallucinations. 

Another exciting direction is to explore the potential of multilingual training and deployment. While the current version was trained exclusively on English data, the starting base model is multilingual. Surprisingly, the multilingual capabilities of the base model persist into the final \textsc{cosmosage} model. This was found by testing \textsc{cosmosage} in Japanese, where it answers coherently. By expanding the training corpus to include cosmological resources in other languages, one could further enhance \textsc{cosmosage}'s multilingual capabilities and further democratize access to cosmological information.

Finally, as the field of LLMs continues to evolve at a rapid pace, it is crucial to stay up to date with the latest advancements and incorporate them into \textsc{cosmosage}'s development. By providing the source code and weights openly, the hope is that this work can be advanced by anyone in the community.

\section{Conclusion}

\textsc{cosmosage} is a natural-language assistant specifically tailored for cosmology. Through continued pretraining on a corpus of cosmology literature and subsequent fine-tuning on a large dataset of synthetically generated question-answer pairs, \textsc{cosmosage} demonstrates an understanding of cosmological concepts and surpasses general-purpose LLMs in this domain.
While quantitative benchmarking remains a challenge due to the lack of standardized tests for specialized LLMs, qualitative evaluation through practical use cases highlights \textsc{cosmosage}'s superior performance in answering cosmology-related questions.
The model's development and deployment, including public availability of the source code, trained weights, and an instance hosted on the Helmholtz AI Blablador inference server, have created a useful assistant that can benefit researchers, educators, students, and anyone interested in a better understanding of the cosmos. 

\section*{Acknowledgements}

This work was made possible by the Center for Computational Astrophysics, National Astronomical Observatory, Japan (CfCA, NAOJ). 
Tijmen de Haan was supported by World Premier International Research Center Initiative (WPI), MEXT, Japan.
Many thanks to Alex Strube for making \textsc{cosmosage} available on the Blablador inference service\secondreply{, and to Cassie Reuter and Joshua Montgomery for acting as independent evaluators.}



%% References


\bibliographystyle{elsarticle-num}
\bibliography{cosmosage}






% \end{thebibliography}

\end{document}